\documentclass{article}
\usepackage{graphicx}

\title{The Effect of Corotation on the Radial Gradient of
Metallicity of Spiral Galaxies \thanks{To appear in the 
Proceedings of IAU Symposium 265 held at Rio de Janeiro, 2009,
K. Cunha, M. Spite \& B. Barbuy, eds. }}

\author{S. Scarano Jr.(1,2), J.R.D. L\'epine(1) \\
(1) Instituto de Astronomia, Geof\'{\i}sica e
Ci\^encias Atmosf\'ericas,\\ Universidade de S\~ao Paulo, Brazil,\\
(2) Southern Astrophysical Research Telescope (SOAR), Chile}

\begin{document}

\maketitle

\begin{abstract}
The corotation radius in a spiral galaxy is the radius where the spiral
pattern speed has the same velocity of the rotation curve. By
compiling results from the literature for 20 spiral galaxies
we verified a strong correlation between the radius of the minima
or inflections of the metallicity distribution  and the
corotation radius.\\

\end{abstract}
\section{Introduction}

The  gradients of metallicity observed in disk galaxies provide
important constraints to the models of the chemical
evolution of these objects. It is frequently observed, in a first
approximation, that spiral galaxies have a  declining gradient of
metallicity. However, it is not rare to observe galaxies with
a clear change in the slope of their gradient of metallicity (eg.
Zaritsky et al., 1994). Considering that spiral arms
are the main contributors to the formation of massive stars,
which in turn are responsible for the metal enrichment of the
interstellar medium, Mishurov et al. (2002) proposed a
simple model for the chemical evolution of spiral galaxies that
effectively introduces the role of the spiral arms in the star
formation rate. This model is able to
explain the plateau observed in the metallicity distribution of our
galaxy, which would be a consequence of the  minimum expected in the
metallicity distribution produced at  corotation. In this work we
compiled a set of spiral galaxies for which the rotation curves, the
gradients of metallicity and the corotation radii (or the spiral
pattern speeds) have been evaluated in the literature, to verify if
the variations in their metallicity gradient are also related to
the corotation.

\section{Sample and Data Analysis}

There are many works on  gradients of metallicity and on the
rotation curves of spiral galaxies, but the corotation radii or the
spiral pattern speeds were only estimated  for a few galaxies. The
observational data for 20 spiral galaxies presented here are
distributed over 108 references compiled and homogenized by
Scarano Jr. (2008). We adopted the Oxygen abundances
observed in H~{\sc ii} regions to evaluate the metallicity
distribution. Distances inside the plane of each galaxy were
recalculated to a same  reference base. All methods used to determine
the corotation were considered equivalent. The changes in the slope
of the metallicity distribution were calculated from a 4th order
polynomial fitted to the observational data, which is enough to
reproduce the expected radial declining behavior of the
metallicities and admits a minimum inside it. Minima and inflections
were determined using the derivatives of the fitted curve. The
uncertainties were calculated by propagating the errors in the
fitted parameters.

\section{Results and Conclusions}

Plotting the minimum or the inflection radius $R_{dZ}$, calculated
with the procedure described above, against the median corotation radius
$R_{CR}$, determined from the literature for each galaxy, and doing
the same thing with the angular velocity $\Omega_{dZ}$, associated
to $R_{dZ}$, against the correspondent spiral pattern speed
$\Omega_{p}$, we obtained the correlations shown in Figure 1
for all galaxies in our sample.

\begin{figure}[b]
\begin{center}
 \includegraphics[width=4.7in]{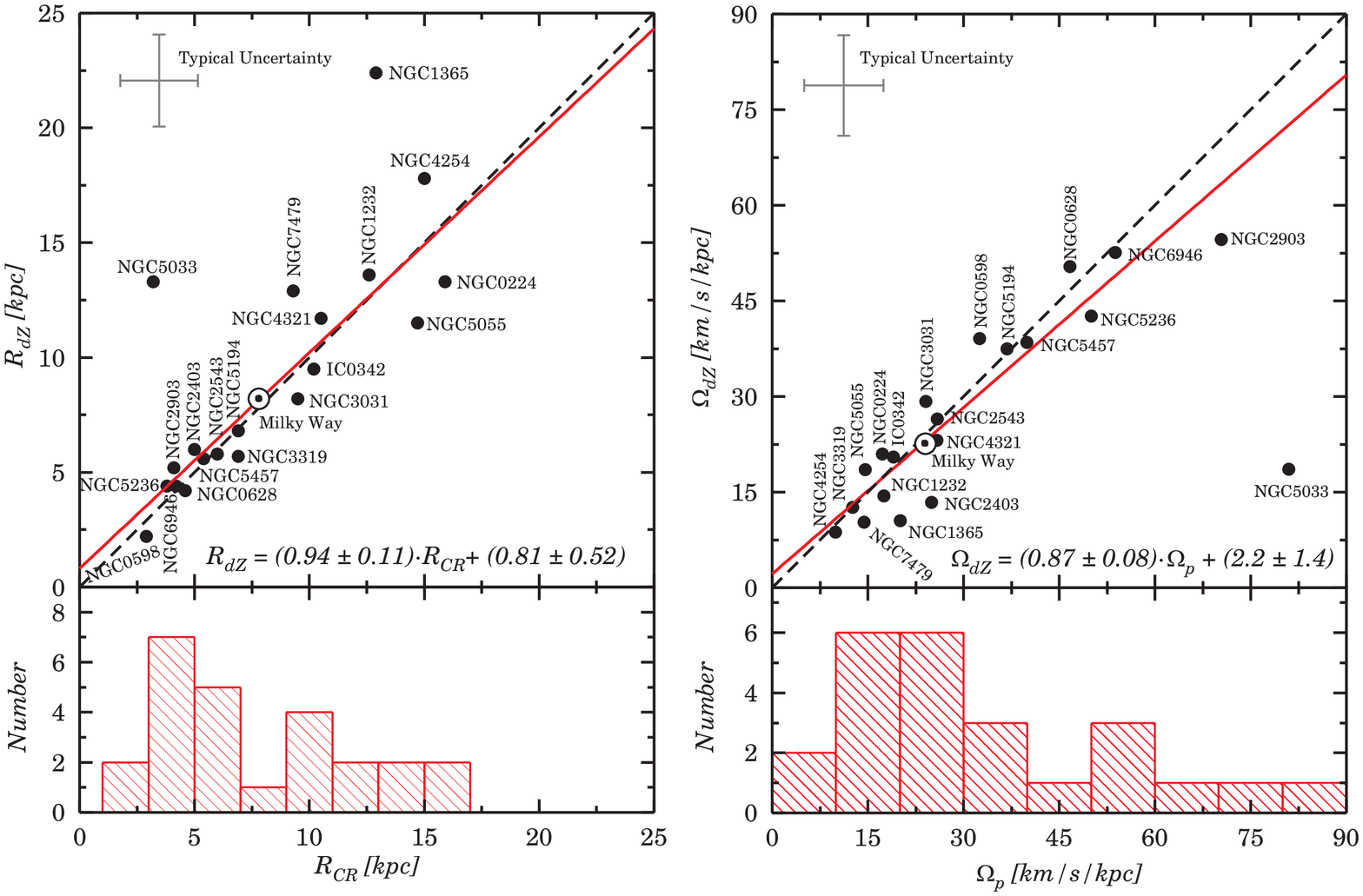}
 \caption{Correlation between the minima or inflection radii in the metallicity
  distribution and the corotation radii (left) and the equivalent graph for
  the angular velocities associated to breaks in the gradients of metallicity
  and the velocities of spiral pattern speed of all sampled galaxies (right).
  The Sun symbol represents the results for our galaxy. The histograms at the
  bottom of each graph show the statistics (number of points per bin) of the data.}
   \label{fig1}
\end{center}
\end{figure}

In this figure the continuous lines are the linear regressions using
robust statistics methods, which show the strong correlation between
the radii of the breaks in the metallicity gradients and  of
corotation. The fitted lines are very near to the one-to-one
correspondence represented by the dashed lines. Concerning other
interesting relations discovered in the present study, the
$\Omega_{p}$ distribution shows a peak around 20 km/s/kpc, revealing
that our galaxy is a typical one in this aspect. A surprising
correlation between the corotation radius and the Elmegreen spiral
arm class was also found, and when it is applied to our Galaxy, it
results in a spiral pattern similar to that revealed by recent
observations from the Spitzer telescope (Churchwell et al.,
2009).

\section{References}

\begin{itemize}

\item[]{Churchwell et al., 2009, PASP 121, 213}

\item[]{Mishurov, Y.N., L{\'e}pine, J.R.D. and Acharova, I.A., 2002, ApJ 571, 113} 

\item[]{Scarano Jr, S., 2008, PhD thesis, Universidade de Sao Paulo.}

\item[]{Zaritsky, D., Kennicutt, Jr., R.C. and Huchra, J.P., 1994, ApJ 420, 87Z}

\end{itemize}

\end{document}